\documentclass[12pt,a4paper]{article}
\usepackage{fullpage,amsmath,amssymb,amsfonts,epsfig,setspace}

\psfigdriver{dvips}

\begin{document}

\title{\begin{bfseries}{Complexity, Collective Effects and Modelling of Ecosystems: formation, function and stability.}
\end{bfseries}}

\author{Henrik Jeldtoft Jensen\footnotemark[1] \footnotemark[2] \footnotemark[3] and Elsa Arcaute\footnotemark[1] \footnotemark[4]}

\maketitle

\fnsymbol{footnote}
\footnotetext[1]{Institute for Mathematical Sciences, Imperial College London, 53 Prince's Gate, South Kensington campus, SW7 2PG, UK.}
\footnotetext[2]{Department of Mathematics, Imperial College London,  South Kensington campus,  London SW7 2AZ, UK.}
\footnotetext[3]{Author for correspondence (h.jensen@imperial.ac.uk; www.ma.ic.ac.uk/$^\sim$hjjens)}
\footnotetext[4]{Department of Physics, Imperial College London, South Kensington campus, SW7 2BW, UK.}

\begin{abstract}
We discuss the relevance of studying ecology within the framework of Complexity Science from a statistical mechanics approach. Ecology is concerned with understanding how systems level properties emerge out of the multitude of interactions amongst large numbers of components, leading to ecosystems that possess the prototypical characteristics of complex systems. We argue that statistical mechanics is at present the best methodology available to obtain a quantitative description of complex systems, and that ecology is in urgent need of ``integrative'' approaches that are quantitative and non-stationary. We describe examples where combining statistical mechanics and ecology has led to improved ecological modelling and, at the same time, broadened the scope of statistical mechanics.  
\end{abstract}


\section{Introduction} \label{sec.introduction}
Does the concept of ``Complexity'' bear any specific meaning or is it just synonymous with complicated and yet not comprehended phenomena? We will argue that it is possible and useful to use the term ``Complexity Science'' in a specific and reasonably well defined way. It is useful because a number of common trends and implications become clear when a phenomenon is classified as part of ``Complexity Science''. The science of complexity emphasises the interactions between components. It stresses that components, most often, are heterogeneous and evolve in time. Complexity is concerned with the emergent properties at systems level originating from the underlying multitude of microscopic interactions. 

In an attempt to make our discussion more clear we will immediately describe the way we use some terms central to our exposition. We hurry to stress that these descriptions are not meant to be exhaustive final philosophical definitions, but rather intended to lower the risk of misunderstanding when we deal with terms frequently used to mean different things by different people. And now our specifications.
\begin{itemize}
\item[]{\bf Complex Systems} consist of a large number of interacting components. The interactions give rise to emergent hierarchical structures. The components of the system and properties at systems level typically change with time. A complex system is inherently open and its boundaries often a matter of convention.
\item[]{\bf Statistical Mechanics} seeks to understand how properties at systems level emerge from the level of the system-components and their interactions. This often involves the application of probability theory, and a number of mathematical techniques. Throughout, we draw a distinction between statistical mechanics and statistical physics. The latter is mainly concerned with the microscopic foundation of thermodynamics and, e.g., phenomena such as phase transitions and superconductivity.
We consider here statistical mechanics as a mathematical methodology, which can be applied to many different sciences including economics, population biology and sociology, to name a few.    
\end{itemize}


Statistical mechanics is a powerful transdisciplinary methodology for the study of emergent phenomena at a macroscopic level caused by the many interactions taking place at a microscopic level. It provides a framework within which it is possible to encapsulate the myriad of degrees of freedom of a system at a microscopic level, into just a few degrees of freedom at a macroscopic level. 
In its current form statistical mechanics does not hold all the answers for all the complex systems, however, we argue that it is at present the best methodology available to obtain a quantitative description of complex systems. By systematically applying it to fields outside its traditional range of application in physics, statistical mechanics can be developed further, in addition to simultaneously contributing to the understanding of those fields, such as ecology. The importance of this feedback loop cannot be overestimated. It can also provide a starting point for the possible development of new mathematical techniques. 
 
Along these lines, the research programme in search for the ``laws'' of ecosystems described by J{\o}rgensen and collaborators \cite{NewEcology}, looks into finding a rigorous set of laws that govern the dynamics at the macro-level. This is a first attempt into establishing a methodology for ecological complexity. At the moment the analysis is mainly qualitative, and we suggest that the second step towards that goal would be the implementation of techniques from statistical mechanics, in order to obtain a rigorous mathematical formalism and modelling.

The paper is organised as follows. In the next section, we will for concreteness illustrate our arguments by briefly describing a complexity inspired model of evolutionary ecology called the Tangled Nature model. This will allow us to demonstrate how macro-evolution can be modelled as emerging from the interacting micro-evolution consisting of individual organisms influencing each other and undergoing reproduction which is prone to mutation. We will discuss feedback, emergence, network structures, and the intermittent temporal mode of macro-evolution in contrast to the steady smooth pace of dynamics at the level of individuals. We will also briefly touch on the modelling of ecological observables such as the Species Abundance Distribution, Species Area Laws and the relationship between interaction and diversity. 

In the Discussion Section we will mention two examples of research presented at the Symposium on  {\it Complexity, Collective Effects and Modelling of Ecosystems: formation, function and stability} at the Beijing Eco Summit 2007. 
These examples illustrate how a complexity science viewpoint may shape the approach of ecological research projects. 
The first example is John Crawford's contribution on ``The Self-organisation of life in Earth''. This work looks at the soil-microbe system, and at the development of models on evolutionary ecology, that can be applied to this dynamical system.
The second example is C{\'e}dric Gaucherel's work on ``Theoretical analysis of dynamic patchy landscapes'', which looks at landscape models constructed within the framework of statistical mechanics.

\section{Tangled Nature Model}\label{sec.TaNa}
\subsection{Description of model}
The Tangled Nature model is defined at the level of interacting individuals. It is an attempt to identify possible simple mechanisms behind the myriad of complicated interactions, feedback loops, contingencies, etc., as one moves from the short time reproductive dynamics at the level of individuals, to the long time systems level behaviour. The strategy is to keep the model sufficiently simple to enable analysis and to pinpoint the details or assumptions in the model that are responsible for the specific behaviour at the systems level. One major concern of the model has been to understand how the smooth continuous pace of the reproductive dynamics at the level of individuals, can lead to intermittent or punctuated dynamics at the level of high taxonomic structures. To be able to address such issues, the model considers individuals as represented by a single sequence with individual number $\alpha$, denoted by ${\bf S}^\alpha= (S_1^\alpha,S_2^\alpha,...,S_L^\alpha)$ belonging to a sequence space ${\cal{S}}$, where all $S_i^\alpha=\pm1$. These sequences undergo simple reproduction during which a given sequence duplicates itself, and while this happens, components of the sequence may mutate, represented by the off-spring having a different sign from the mother, i.e. $S_i^\gamma=-S_i^\alpha$, where $\gamma$ denotes the index for the daughter, and $\alpha$ the one for the mother. The aim of the model is to understand the macro-dynamics emerging at the systems level. This is done by analysing the dynamics of the occupancy in this sequence or type space. Taxonomic structures, such as species formation, emerge as aggregations in the density of occupied sites $n({\bf S},t)$ in the type space. This is very much in accordance with Mallet's definition of species \cite{mall:spec}. A species will be identified as a local peak in the density $n({\bf S},t)$, and species formation will correspond, e.g., to the splitting of such a peak into two peaks. Macroscopic ecological measures such as Species Abundance Distributions are derived from the structure of $n({\bf S},t)$. 

Let us now sketch the mathematical details of the model. For in depth studies of the model, please refer to \cite{TaNaBasic,TaNaNetwork,TaNaQuasi,TaNaTimeDep}. The size of the type space is set by the length, $L$, of the sequences; a typical value used is $L=20$ leading to about one million different genotypes. The sites in the genome space are supposed to represent all possible ways of constructing a 'genome'. Many sequences may not correspond to viable organisms. The viability of a genotype is determined by the evolutionary dynamics. All possible sequences are available for evolution to select from. We will see that a natural species concept arises from the dynamics, in which each species is separated in genotype space. 

The system consists of $N(t)$ individuals, and a time step consists of {\em one} annihilation attempt followed by {\em one} reproduction attempt. A reproduction event is successful with varying probability $p_{off}$, defined below, and an annihilation attempt is successful with constant probability $p_{kill}$. The killing probability is considered a constant independent of type for simplicity. It would obviously be more realistic to let $p_{kill}$ depend on the type of the individual considered. However, this does not change the overall behaviour at systems level. One generation consists of $N(t)/p_{kill}$ time steps, which is the time taken (on average) to kill all currently living individuals. The dynamics leads to a population size which remains nearly constant on short timescales. The individuality of the specific types, or sequences, is given by their ability to reproduce. Since we are interested in the collective, or complexity, aspects of evolution, the Tangled Nature model stresses the mutual influence amongst different types of organisms. This is done by assuming that each individual of type ${\bf S}$ is able to reproduce, when selected for reproduction, with a probability $p_{off}({\bf S},t)$ that depends on the sequence ${\bf S}$ and the configuration of other types in the type space. The reproduction probability, $p_{off}$, is determined by a weight
function $H({\bf S}^\alpha,t)$:  
\begin{equation}
H({\bf S}^\alpha,t)={\frac{c}{N(t)}} \left( \sum_{{\bf S}\in{\cal S}} 
{\bf J}({\bf S}^\alpha,{\bf S})   n({\bf S},t) \right)
- \mu N(t),
\label{Hamilton2}\index{weight function}
\end{equation}
where $c$ controls the strength of the interaction (large $c$ means a large interaction),
 $N(t)$ is  the total number of individuals at time $t$, 
the sum is over the $2^L$ locations in ${\cal S}$ and $n({\bf S},t)$ is 
the number of individuals (or occupancy) at position ${\bf S}$. 
Two positions ${\bf S}^a$ and ${\bf S}^b$ in genome space are coupled with fixed but random 
strength ${\bf J}({\bf S}^a,{\bf S}^b)$ which can
 be either positive, negative or zero. 
This link exists (in both directions) with probability $\theta$, i.e. $\theta$ is
simply the probability that any two sites are interacting.  If the link exists, then ${\bf J}({\bf S}^a,{\bf S}^b)$ 
and ${\bf J}({\bf S}^b,{\bf S}^a)$  are both generated randomly and independently, and such that they belong to $(-1,1)$.
To study the effects of interactions \textit{between} species, we exclude
self-interaction so that ${\bf J}({\bf S}^a,{\bf S}^a) =0$. 

The conditions of the physical environment are simplistically described by the term $\mu N(t)$ in Eq. (\ref{Hamilton2}),  where $\mu$ determines the average sustainable total population size, i.e. the carrying capacity of the environment. This is an example of how the question of the openness and ``surroundings'' of ecosystems arises in a natural way in the present statistical mechanics like formalism. An increase in $\mu$ corresponds to harsher physical conditions. We use asexual reproduction consisting of one individual being replaced by two copies mimicking the process of binary fission seen in bacteria. We allow for mutations in the following way: with probability $p_{mut}$ per gene we perform a change of sign $S_i^\alpha \rightarrow - S_i^\alpha$ during reproduction. Successful reproduction occurs with a probability per unit time, $p_{off}({\bf S}^\alpha,t)\in[0,1]$, given by
\begin{equation}
p_{off}({\bf S}^\alpha,t)={\frac{ \exp[H({\bf S}^\alpha,t)]}
{1+\exp[H({\bf S}^\alpha,t)]}}.
\label{p_off}
\end{equation}
This function is chosen for convenience, since the specific functional form has no effect on the dynamics of the model - any smoothly increasing function that maps $H({\bf S}^\alpha,t)$ to the interval $[0,1]$ will do. Let us mention that this basic quantity is deliberately taken by the Tangled Nature model to be a context dependent reproduction probability rather than a fitness function. One reason why this is done is to try to avoid some of the dangers and subtleties inherent to the fitness concept \cite{ConfusFitness}. 

Eq. (\ref{Hamilton2}) can be understood as the interaction of an individual with all the
others, with a term $\mu N$ which determines the 
total population and controls fluctuations.  The interaction strength $c$
gives the magnitude of the total interaction.  We can tune the effective `resource'
density (and hence the population density) with the parameters $\mu$.  The total population 
remains approximately constant over ecological timescales (and actually increases over evolutionary timescales). Setting self-interaction to zero is equivalent to considering that all types interact equally with their own species (one can rescale $p_{kill}$ and $\mu$ to accommodate this). This constraint is imposed in order to focus on the effect of interactions between different types. To study the relation between diversity and the strength of the interactions, there is a version of the model in which different strengths of intra-specific interactions are included \cite{DivInt}. Obviously it is impossible to design the details of the interaction matrix $J({\bf S}^a,{\bf S}^b)$ in a realistic way. What can be accomplished is to study qualitative questions, such as what is the effect of having very few interaction links between sequences compared with many interaction links \cite{TaNaNetwork}. Or one can address the effect of correlations in the allowed interactions \cite{lair05:tang}?

After a short transient period the initial state becomes irrelevant. There are two very different initial conditions that consist in placing the entire population at time zero in: i) one position in type space, or ii) on random positions, i.e. a random collection of initial types. Since both configurations are badly adapted to the interaction matrix $J({\bf S}^a,{\bf S}^b)$, in both cases the population will typically collapse to one single position in type space. Eventually the population size will have decreased enough to make the $-\mu N(t)$ term sufficiently small to allow $p_{off}$ to grow to a value that ensures a non-vanishing reproduction rate. When this happens the population will, due to mutations, start to spread out from its initial position into the surrounding genotype space. And as this happens, natural selection will ensure that only certain configurations of occupied sites are viable. These are configurations for which the mutual interactions between the types lead to off-spring probabilities that, for a significant part of the occupied types, are able to balance the killing probabilities, i.e.  $p_{off}({\bf S},t)=p_{kill}$ for some set of types ${\bf S}$.
        \begin{figure}[h]
        \centering\noindent
        \includegraphics[width=16cm]{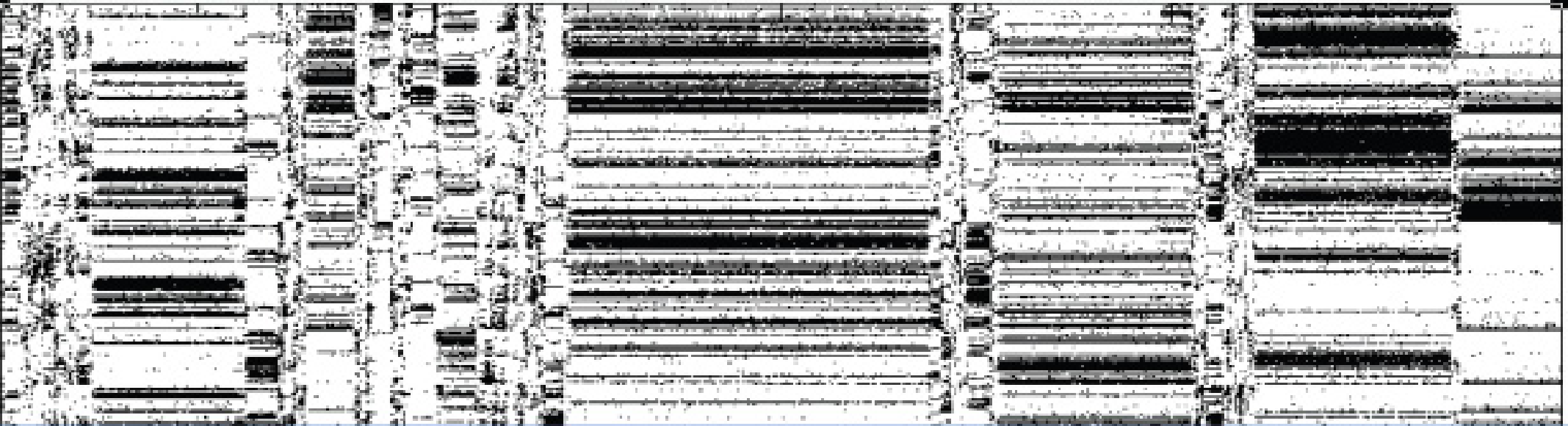}
      \caption{Intermittent evolution of the occupancy in type space. Time, measured in generations, is along the x-axis. The about $10^6$ different types are labelled up along the y-axis. Whenever a type is occupied a dot is placed at its label. Long stretches of parallel lines indicate epochs during which the main composition in type space remains essentially the same. Figure courtesy of Matt Hall.}
      \label{Intermittency}
        \end{figure}

The dynamics in type space is characterised by a two-phase switching, consisting of long periods of relatively stable configurations (quasi-Evolutionary Stable Strategies or q-ESSs) (Fig. \ref{Intermittency}) interrupted by brief spells of reorganisation of occupancy called transitions. Transition periods are terminated when a new q-ESS is found, as discussed in \cite{TaNaBasic}. The intermittent macro-dynamics is not in a stationary state.  When one considers very many realisations of the dynamics it turns out that the transition rate between q-ESS decreases with the age of the system \cite{TaNaTimeDep}. This happens because selection is able to pick out configurations in type space that tend to possess more beneficial links (i.e. positive ${\bf J}({\bf S}^a,{\bf S}^b)$ bonds) than is the case between a randomly selected set of types. We consider this directedness of the long time systems level dynamics to prototypical of complex systems \cite{Anderson04}.

\subsection{Emergent Ecological Measures}
As we move from the level of individuals to the systems level, `species' can be well defined as the highly occupied genotype points called `wildtypes', which are separated in genotype space.  Each wildtype is surrounded by a `cloud' of mutant genotypes with low occupancy.  Thus we can take a natural definition of diversity: the number of wildtypes in the system. It is interesting to study how the Species Abundance Distribution (SAD) depends on the assumed properties of the interaction matrix ${\bf J}({\bf S}^a,{\bf S}^b)$. It was found by \cite{TaNaNetwork} that the often observed log-normal shape of the SAD is reproduced by the evolutionary dynamics of the Tangled Nature model under certain conditions. Namely, when each type is potentially able to interact with a large number of other types. In this case the adapted configurations consist of populations of species that form one large interconnected cluster and the SAD evolves with time towards a log-normal like form. If ${\bf J}({\bf S}^a,{\bf S}^b)$ only allows a type to interact with few other types (i.e. few non-zero elements in the $J$ matrix), the population in the type space splits up into separate groups and the SAD doesn't develop a form resembling a log-normal distribution. Considered from this perspective the SAD might be thought of as containing information about the properties of the network of all possible interactions between organisms.

Let us focus on the properties of the network of interactions\footnote{The nodes of the network under consideration, consist of occupied positions in type space. There is an edge between two nodes ${\bf S}^a$ and ${\bf S}^b$ if the two types interact, i.e. if ${\bf J}({\bf S}^a,{\bf S}^b)$ or ${\bf J}({\bf S}^b,{\bf S}^a)$ is non-zero.} of extant species. The evolutionary dynamics performs a collective adaptation on the co-existing types in type space. As selection and adaptation act generation after generation, a subset of sites in type space becomes occupied. This subset is selected such that the mutual interactions allow each of the extant species to counter balance the depletion of its population, caused by death ($p_{kill}$) and mutations ($p_{mut}$), by a sufficiently large off-spring production ($p_{off}({\bf S},t)$). The network of interactions between these co-existing types possesses some interesting emergent properties. The typical coupling strength between extant types is more mutualistic than the coupling between arbitrary types ${\bf S}^a$ and ${\bf S}^b$, chosen at random in type space irrespectively of the types being extant or not \cite{TaNaNetwork,lair05:tang}. This effect is significantly bigger when the coupling matrix ${\bf J}({\bf S}^a,{\bf S}^b)$ is correlated for sites ${\bf S}^a$ and ${\bf S}^b$ that reside in the same vicinity of type space. A correlated coupling matrix is more realistic, since it corresponds to assuming that similar organisms have a certain similarity in the way they interact with the surrounding ecosystem. The degree distribution of the network of extant species is sensitive to the amount of correlations imposed on type space. When correlations are present, we typically observe exponential degree distributions of the network of interactions between extant types. In contrast, uncorrelated interaction matrices lead to binomial degree distributions, as it is observed in networks where edges are placed at random. This example indicates how some properties at systems level may be caused by generic mechanisms for emergent collective behaviour. 

A similar situation is encountered when the qualitative behaviour of spatial properties is investigated. Spatial aspects are obviously of the greatest ecological importance. A simple quantity to start out with is the Species Area Relation (SAR). By placing a copy of the Tangled Nature model on each site of a two dimensional lattice, one can make a simplistic model combining evolutionary dynamics with spatial dispersion. Such a model was studied in \cite{laws05:spec} and a power-law SAR is observed. The evolutionary dynamics produces a high degree of spatial diversity even when the same type space is placed on each site of the spatial lattice. 

Complexity science stresses that the {\em interaction} between the components is responsible for the emergent properties at systems level. 
Sometimes for tractability reasons, models might oversimplify the components compared with reality, and still it happens that such simple models are able to capture certain qualitative aspects.
An attempt in this direction was made in \cite{DivInt}, where the relationship between interactions amongst different types and the diversity of types was discussed. The inspiration behind this study came from molecular evolution experiments on {\it E. coli}, in which the relation between fitness plasticity and diversification was addressed \cite{DiverseEcoli}. The model used a version of the Tangled Nature model in which each type was assigned an amount of self-interaction. This was achieved by supplementing the weight function in Eq. (\ref{Hamilton2}) by an extra term proportional to an intrinsic fitness $E({\bf S}^\alpha)$. It was found that diversity rapidly increased when the typical interaction strength, set by the parameter $c$ in Eq. (\ref{Hamilton2}), exceeded a certain value, determined by the properties of the intrinsic fitness.      

\section{Discussion}
We have proposed  that complexity science can be seen as a coordinated attempt to understand how emergent collective behaviour at systems level arises due to the multitude of interactions between the components. From this perspective we have argued that complexity science offers a particularly relevant approach to ecology. Above we tried to illustrate our point of view with some theoretical examples taken from the Tangled Nature model's study of evolutionary ecology. Let us now conclude with a couple of examples that are closely related to observations, where we believe that the complexity science's perspective has made a difference. \\


Our first example is the soil-microbe system. In the study of such systems, there is an urgent need to develop models on evolutionary ecology that integrate function and diversity, and that are dynamical \cite{Crawford2}, at present many of the current models for soil are static.  
The soil-microbe system is an extremely rich and intricate system that has not yet been fully understood, and moreover is of great relevance to agriculture, waste management and the water industry to mention a few \cite{Crawford3}. Crawford and collaborators believe that any model describing the system should integrate biochemistry and biophysics, since from the interactions between the biotic and abiotic factors, the structure, the functionality and the dynamical behaviour of the soil emerge. 

The soil-microbe complex is a self-organised system capable to adapt, therefore, within the many different approaches that currently exist \cite{Crawford2}, they suggest the development of models of evolutionary ecology that have the same aspects as those described by the Tangled Nature model: evolutionary processes, population dynamics, feedback loops, interactions, etc. In addition, the ecosystem functioning needs to be included, since it is the relation between diversity and function that is mostly ignored in other soil models. Experimental techniques to measure this connection need therefore to be developed. They argue that the ecosystem-level behaviour of the soil-microbe system is the outcome of the behaviour at the organism's level, which is natural if the system is classified as a complex system. In addition, this viewpoint stresses the importance to include evolutionary processes when looking at ecology, since these are crucial for the understanding of ecological function. 
For example, in the soil-microbe system, the activity of the microbes changes the structure of the soil by affecting its rates of oxygen diffusion and porosity, while the substrates in the soil affect the activity of the microbes.

Crawford and collaborators have identified important properties of soil systems by applying statistical mechanics to their research \cite{Crawford1}, however, they also recognise the need of an extended methodology, where interdisciplinarity is crucial. They emphasise that there is no unique discipline that is able by itself to understand the soil \cite{Crawford3}.\\


Our second example is the modelling of landscapes. Models explaining and being able to predict the shape of landscapes are extremely important to prove ecological hypotheses, and for the implementation and development of market and land planning policies. Gaucherel and collaborators \cite{Gaucherel3}, argue that in many systems, the most relevant factors causing the dynamical changes of landscapes are human driven. For this reason they urge for the development of models that integrate biophysical and socio-economical factors. They propose a generic modelling platform: ``L1'', that can be used to look at the patterns resulting from specific processes. This can therefore be applied as a tool to assess environmental policies and technological implementations at different landscape scales. The platform simulates the dynamical evolution of a landscape as a result of the feedbacks and interactions between the elements composing the landscape. 
Following the methodology of statistical mechanics, the objects are modelled by introducing only the relevant aspects that give rise to such emergent structures, and not by parametrising all of their degrees of freedom.

In their approach, Gaucherel and collaborators stress the crucial role of feedback and scales in ecological systems. On the one hand, the landscape is an emergent structure, product of the interactions between the many different components, see for example the role of farm systems in human driven landscapes \cite{Gaucherel3}.
On the other, their models give rise to hierarchical structures that feed back into the system. For example, the landscape itself determines important aspects of habitats and ecosystems at different scales.  
This is illustrated in \cite{Gaucherel1}, where the authors look at the relationship between the characteristics of the habitat given by the landscape, and particular characteristics of the inhabitant species, such as their spatial distribution, their morphology, etc. Using their framework, one can investigate at different locations, times and scales, the ecological relationships. They apply their model to look at the link between agricultural activities, landscape shape and some characteristics of carabid beetles, such as their abundance and their body size. They proved that some correlations are only valid for certain specific locations and scales, contrary to what was believed using other techniques unable to give spatial and scaling precision.

An explicit outline of how the methodology from statistical mechanics is implemented in this approach to landscape modelling can be found in \cite{Gaucherel2}. There the authors construct a neutral model for patchy landscapes using the Gibbs process to describe the interactions between the different components of the landscape. A neutral model is a model that simulates the landscape properties and patterns that are not the outcome of a particular ecological process. Therefore, using these models a distinction can be drawn between structures caused by random processes and those obtained through real processes. In addition, such models give rise to virtual landscapes that can be used to study real mosaics, such as forest landscapes \cite{Gaucherel2}. 
For this or other specific applications, the neutral model is implemented in the L1 platform mentioned above, where the particularities of the system can be entered.\\

The project to establish a systems perspective on ecology as laid out in the book {\em A New Ecology} by J{\o}rgensen and collaborators in \cite{NewEcology} can, in our opinion, be seen as a prototypical example of the objectives aimed at when taking inspiration from the methodology of statistical mechanics and applying it to complex systems. J{\o}rgensen et al. argue that laws at the  emergent systems level may exist and the authors suggest a list of laws they believe ecosystems obey. They make clear that it is a grand task to identify these laws and that their list is to be thought of as a starting point. If we embed these laws within the statistical mechanics framework, they describe the expected properties of a complex system. 
For example, J{\o}gensen et al. mention that ecosystems have openness, connectivity, complex dynamics and that their dynamics is directed. 
These properties are totally in agreement with the features of complex systems highlighted by applying methods from statistical mechanics to their analysis. 

In addition, within the Tangled Nature model, interconnectedness and interaction become a focal point of the description when one thinks in terms of the emergent networks of interactions between extant species. Furthermore, within the mathematical formulation of dynamical systems, there is a term encoding the carrying capacity and resources, which represents an open system. Ecosystems are therefore correctly taken into account as open within this formalism.
The Tangled Nature model was formulated by including what appears to be minimal assumptions for the dynamics; namely reproduction prone to mutations and livelihoods of the individual types that are influenced by other coexisting types. As a result the model produces a slowly adapting non-stationary directed dynamics at the macroscopic systems level. This is certainly in agreement with properties of ecosystems encoded in the form of a law by J{\o}rgensen et al. and, moreover, it appears to be in agreement with records of macro evolution \cite{Newman99c}. The non-stationary directional nature of the Tangled Nature model has been suggested to be an example of generic properties of complex systems dynamics as observed in a number of very diverse phenomena by Anderson et al. \cite{Anderson04}.  Here it was suggested that the directional gradual relaxation can be viewed as a release of a generalised strain originating in starting from a badly adjusted initial configuration. The Tangled Nature model suggests that in the case of ecosystems, selection and adaptation manage to direct the dynamics towards a selections of species better adjusted to  coexist. \\
 
In conclusion, it is useful to identify systems under the label ``Complex Systems'', since this indicates that the machinery of statistical mechanics can be applied to try to describe the system's dynamics and evolution. 
In ecology, there is an urgent need for ``integrative'' approaches that are non-stationary, and statistical mechanics can provide an initial mathematical framework,
subject to modification and adaptation as one navigates deeper into the mysteries of complex systems.

\section*{\small{Acknowledgements}}

\small{We wish to thank the participants in the Symposium on {\em Complexity, Collective Effects and Modelling of Ecosystems} for interesting discussions at the Beijing Eco Summit 2007. HJJ is grateful to P Anderson, K. Christensen, S. A. di Collobiano, M. Hall,  S. Laird and D. Lawson for stimulating collaboration.  }


\begin{thebibliography}{10}

\bibitem{NewEcology}
S.E. J{\o}rgensen et~al.
\newblock {\em A New Ecology. {S}ystems Perspective.}
\newblock Elsevier, 2007.

\bibitem{mall:spec}
James Mallet.
\newblock {A} species definition for the {M}odern {S}ynthesis.
\newblock {\em Trends Ecol. Evol.}, 10(7):294--299, 1995.

\bibitem{TaNaBasic}
Kim Christensen, Simone~A. di~Collobiano, Matt Hall, and Henrik~J. Jensen.
\newblock Tangled nature: A model of evolutionary ecology.
\newblock {\em J. Theor. Biol.}, 216:73--84, 2002.

\bibitem{TaNaNetwork}
Paul Anderson and Henrik~Jeldtoft Jensen.
\newblock Network properties, species abundance and evolution in a model of
  evolutionary ecology.
\newblock {\em J. Theor. Biol.}, 232:551--558, 2005.

\bibitem{TaNaQuasi}
Simone~Avogadro di~Collobiano, Kim Christensen, and Henrik~Jeldtoft Jensen.
\newblock The tangled nature model as an evolving quasi-species model.
\newblock {\em J. Phys A}, 36:883--891, 2003.

\bibitem{TaNaTimeDep}
Matt Hall, Kim Christensen, Simone~A. di~Collobiano, and Henrik~Jeldtoft
  Jensen.
\newblock Time-dependent extinction rate and species abundance in a
  tangled-nature model of biological evolution.
\newblock {\em Phys. Rev. E}, 66:011904, 2002.

\bibitem{ConfusFitness}
Andr{\'e} Ariew and R.C. Lewontin.
\newblock The confusion of fitness.
\newblock {\em Brit. J Phil. Sci}, 55:347--363, 2004.

\bibitem{DivInt}
Daniel~John Lawson, Henrik~Jeldtoft Jensen, and Kunihiko Kaneko.
\newblock Diversity as a product of interspecial interactions.
\newblock {\em J. Theor. Biol.}, 243:299--307, 2006.

\bibitem{lair05:tang}
Simon Laird and Henrik~Jeldtoft Jensen.
\newblock {T}he tangled nature model with inheritance and constraint:
  Evolutionary ecology restricted by a conserved resource.
\newblock {\em Ecol. Complexity}, 3:253--262, 2006.

\bibitem{Anderson04}
P.E. Anderson, H.J. Jensen, L.P. Oliveria, and P.~Sibani.
\newblock {\em Complexity}, 10:49--56, 2004.

\bibitem{laws05:spec}
Daniel~John Lawson and Henrik~Jeldtoft Jensen.
\newblock {T}he species area relationship and evolution.
\newblock {\em J. Theor. Biol.}, 241:590--600, 2006.

\bibitem{DiverseEcoli}
Akiko Kashiwagi, Wataru Noumachi, Masato Katsuno, Mohammad~T. Alam, Itaru
  Urabe, and Tetsuya Yomo.
\newblock Plasticity of fitness and diversification process during an
  experimental molecular evolution.
\newblock {\em J. Molec. Evol.}, 52:502--509, 2001.

\bibitem{Crawford2}
J.W. Crawford, J.A. Harris, K.~Ritz, and I.M. Young.
\newblock Towards an evolutionary ecology of life in soil.
\newblock {\em Trends Ecol. Evol.}, 20:81--87, 2005.

\bibitem{Crawford3}
I.~M. Young and J.~W. Crawford.
\newblock Interactions and self-organization in the soil-microbe complex.
\newblock {\em Science}, 304:1634--1637, 2004.

\bibitem{Crawford1}
X.X. Zhang, S.N. Johnson, P.J. Gregory, J.W. Crawford, I.M. Young, P.J. Murray,
  and S.C. Jarvis.
\newblock Modelling the movement and survival of the root-feeding clover
  weevil, sitona lepidus, in the root-zone of white clover.
\newblock {\em Ecol. Modelling}, 190:133--146, 2006.

\bibitem{Gaucherel3}
C.~Gaucherel, N.~Giboire, V.~Viaud, T.~Houet, J.~Baudry, and F.~Burel.
\newblock A domain-specific language for patchy landscape modelling: The
  brittany agricultural mosaic as a case study.
\newblock {\em Ecol. Modelling}, 194:233--243, 2006.

\bibitem{Gaucherel1}
C.~Gaucherel, F.~Burel, and J.~Baudry.
\newblock Multiscale and surface pattern analysis of the effect of landscape
  pattern on carabid beetles distribution.
\newblock {\em Ecol. Indicators}, 7:598--609, 2007.

\bibitem{Gaucherel2}
C.~Gaucherel, D.~Fleury, D.~Auclair, and P.~Dreyfus.
\newblock Neutral models for patchy landscapes.
\newblock {\em Ecol. Modelling}, 197:159--170, 2006.

\bibitem{Newman99c}
M.~E.~J. Newman and P.~Sibani.
\newblock Extinction, diversity and survivorship of taxa in the fossil record.
\newblock {\em Proc. R. Soc. Lond. B}, 266:1--7, 1999.

\end{thebibliography}
\end{document}